\documentstyle[preprint,aps]{revtex}
\begin{document}
\title{Plane symmetric analogue of NUT space}
\author{M.~Nouri-Zonoz$^{a,b}\thanks{Electronic address:~nouri@theory.ipm.ac.ir}$ and
A.~R.~Tavanfar$^{a}\thanks{Electronic address:~tavanfar@molavi.ut.ac.ir}$ }
\address{$^{a} $ Department of Physics, Tehran University, End of North Karegar St.,
Tehran 14352, Iran.\\
$^{b}$Institute for studies in theoretical physics and
Mathematics, P O Box 19395-5531 Tehran, Iran.}
\maketitle
\begin{abstract}
In this article on the basis of a new definition of spacetime
symmetry, which is in accordance with the symmetry of the
curvature invariants, we investigate exact vacuum solutions of
Einstein field equations corresponding to both static and
stationary {\it plane symmetric} spacetimes using the concepts of
the (1+3)-decomposition or {\it threading} formalism. Demanding
the presence of a plane symmetric gravitomagnetic field we find a
family of two parameter ($m$ and $\ell$) solutions, every member
of which being the plane symmetric analogue of NUT space.
\end{abstract}

\pacs{PACS numbers:~0420J, 0440N}

\section{Introduction}
The $(1+3)$-decomposition (threading) of a spacetime  by a
congruence of timelike curves (observer worldlines) leads to the
following splitting of the spacetime interval element [1];
$$ds^{2}=dT^{2}-dL^{2}\eqno(1)$$
where $dL$ and $dT$ are defined to be {\it the invariant spatial
and temporal length elements} of two nearby events respectively.
They are constructed from the normalized tangent vector $u^{a}=
{\xi^a \over |\xi|}$  to the timelike curves in the following way
\footnote{Note that Latin indices run from 0 to 4 while the Greek
ones from 1 to 3 and throughout we use gravitational units where
c=G=1.};
$$dL^{2}= h_{ab}dx^{a}dx^{b}\eqno(2)$$
$$dT= u_{a}dx^{a}\eqno(3)$$
where
$$h_{ab}= -g_{ab}+u_{a}u_{b}\eqno(4)$$
is called the {\it projection tensor}. Taking $h\equiv |\xi|^2$
and $A_a \equiv -{\xi_a \over |\xi|^2} $, equations (1) and (4)
can also be written in the following alternative forms;
$$ds^{2}=  h(A_a {\rm d}x^a)^2 - h_{ab}{\rm d}x^a
{\rm d}x^b \eqno(5)$$
$$h_{ab}= -g_{ab}+h A_{a}A_{b}$$
Using the preferred coordinate system in which the timelike curves
are parameterized by the coordinate time $x^{0}$ of the comoving
observers\footnote{Hereafter equations written in this preferred
coordinate system are denoted by the sign $\doteq$.};
$$\xi^a \doteq (1,0,0,0)\;\;\;\;\;\ \& \;\;\;\;
A_a \doteq (-1, - {g_{0\alpha}\over g_{00}})\eqno(6)$$ then above
spatial and spacetime distance elements will take the following
forms [2];
$$dL^2 \doteq dl^{2}=\gamma_{\alpha\beta}dx^{\alpha}dx^{\beta}\eqno(7)$$
$$ds^{2}=e^{2\nu}(dx^{0}-A_{\alpha}dx^{\alpha})^{2}-dl^{2}\eqno(8)$$
where
$$e^{2\nu}\equiv g_{00} \;\;\;\;\;\;\;\;\;\;\; A_{\alpha}\doteq -g_{0\alpha}/g_{00}\eqno(9)$$
and
$$\gamma_{\alpha\beta} = (-g_{\alpha\beta} +{g_{0\alpha} g_{0\beta}\over g_{00}}).\eqno(10)$$
Introducing {\it gravitoelectric} \footnote{For Reviews on the
subject of Gravitoelectromagnetism see references [3] and [4].}
and {\it gravitomagnetic} fields [3];
$${{\bf E}_g} = -\nabla {\nu}\eqno(11)$$
$${\bf B}_g = {\rm curl} {\bf A }\eqno(12)$$
one can write the vacuum Einstein equations for stationary
spacetimes \footnote{For stationary spacetimes the preferred
coordinate system  is the one adapted to the congruence of its
timelike killing vector, ${\bf \xi}_t=\partial_t$.} in the
following quasi-Maxwell form [3];
$${\rm div} {\bf B}_g = 0 \eqno(13)$$
$${\rm Curl} {\bf E}_g =0 \eqno(14)$$
$${\rm div} {\bf E}_g=-{1\over 2}e^{2\nu}B_g^{2} + E_g^{2} \eqno(15)$$
$${\rm Curl}(e^{\nu}{\bf B}_g)=2{\bf E}_g \times {e^{\nu}{\bf B}_g} \eqno(16)$$
$$P^{\alpha\beta}=E_g^{\alpha;\beta}+e^{2\nu}(B_g^{\alpha}B_g^{\beta}-{B_g}^{2}\gamma^{\alpha\beta})+
 E_g^{\alpha}E_g^{\beta} \eqno(17)$$
where $P^{\alpha\beta}$ is the three dimensional Ricci tensor
constructed from the metric $\gamma^{\alpha\beta}$. One can show
that the gravitoelectromagnetic fields can be written in the
following covariant forms;
$$(E_g)_b = -{1\over 2} {(\xi_a \xi^a)_{;b}\over |\xi|^2}
=-{1\over 2}{h_{;b}\over h}\eqno(18)$$
$$B_g^b = -{1\over 2} |\xi|
\xi^{a}\varepsilon_a^{b cd} \left[ ({\xi_d \over |\xi|^2})_{;c} -
({\xi_c \over |\xi|^2})_{;d} \right]={1\over 2}\sqrt{h}\xi^a
\varepsilon_a^{bcd}(A_{d;c} - A_{c;d})\eqno(19)$$ One can also
show that the gravitational Lorentz force on a test particle due
to the spacetime curvature is given by [2,3];
$${\bf f} = {m_{0} \over \sqrt{1-v^{2}}}[{{\bf E}_g} +
{\bf v} \times(e^\nu {{\bf B}_g})]\eqno(20)$$ This force deviates
test particles from geodesics of the 'space' and make them follow
the geodesics of spacetime [2,3]. It should be noted that the
above ( gravitoelectromagnetic ) vector fields ${\bf E}_g$, ${\bf
B}_g$, ${\bf A}$ and the 3-dimensional tensor field
$P^{\alpha\beta}$ are defined on a 3-dimensional manifold $\Sigma$
(whose metric is $\gamma_{\alpha\beta}$). For a general spacetime
manifold $\cal M$, $\Sigma$ is defined as the factor space ${\cal
M}\over \sim$, where $\sim$ is the equivalence relation which
brings all the points on each threading curve under one class. In
the case of stationary spacetimes this equivalence relation is
given by the one-dimensional group $G_1$ of transformations
generated by its timelike Killing vector ${\bf \xi}_t$ [5].
Physically this manifold, with the above spatial distance element
$dl$, can be recognized as {\it the observed (3-dimensional) space
of events} in the sense that $dl$ is the spatial distances between
events as measured by the observers on the timelike curves of the
congruence [3,5]. From the mathematical point of view one should
note that this is an abstract 3-manifold whose Riemannian metric
$\gamma_{\alpha\beta}$ does not correspond to any hypersurface as
its natural habitat!. This is in contrast to the
(3+1)-decomposition (or slicing ) of spacetimes [6] where the
3-dimensional metric lives on the spacelike hypersurfaces. In fact
it can be shown that $\Sigma$ has a one-parameter family of
Riemmanian metrics and in this way one can describe spacetimes in
the language of {\it parametric manifolds} [6,7].
\section{ What do we mean by a plane symmetric spacetime?}
Plane symmetric static spacetimes have been found long time ago by
Taub [8]. But using the concepts of 'threading' approach to the
spacetime decomposition
 one can show that finding a metric with a certain symmetry is a
different problem from finding a spacetime with the same symmetry.
A known example is the NUT spacetime [9], although the metric
itself dose not have spherical symmetry but the spacetime really
does i.e. all the curvature invariants of the spacetime are
spherically symmetric [10,3]. The same property have been shown
for the cylindrical symmetry through the so called {\it
cylindrical analogue of NUT space} [11]. This difference is due to
the fact the physical symmetry of a spacetime is the one
associated with its spatial metric $\gamma_{\alpha\beta}$ and its
gravitoelectromagnetic fields ${\bf E}_g $ and  ${\bf B}_g$ and
therefore it may or may not be followed by the  gravitomagnetic
potential ${\bf A}$ [3] (See section III below for a general
derivation ). As a matter of fact under the simultaneous
transformations;
$$A_{\alpha}\rightarrow A^{\prime}_{\alpha}=
A_{\alpha}+ \phi_{,\alpha}(x^\beta)\eqno(21)$$
$$g_{\alpha\beta}\rightarrow g^{\prime}_{\alpha\beta}=
g_{\alpha\beta}+g_{00}(A_{\alpha}\phi_{,\beta} +
A_{\beta}\phi_{,\alpha}+(\phi_{,\alpha})(\phi_{,\beta}))=
g_{\alpha\beta}+g_{00}(A^{\prime}_{\alpha}-
A_{\alpha})(A^{\prime}_{\beta}+A_{\beta})\eqno(22)$$ the spatial
metric $\gamma_{ab}$ and $g_{00}$ are unchanged i.e. metric will
be physically unchanged, except for the changes in its time zero
i.e. [2];
$$x^{0}\rightarrow x^{0}-\phi(x^\alpha)\eqno(23)$$
This is not only true for stationary spacetimes but also for
non-stationary spaces and it can be taken as a hint for studying
the non-stationary spacetimes in the context  of
gravitoelectromagnetism [13].
\section{Physical symmetry of a spacetime : Gauged motion }
The
well known formulation of spacetime symmetries is based on the
concepts of Killing vectors and the Killing equation (or Killing
motion);
$$\pounds_\xi g_{a b} = 0 \eqno(24)$$
which is obtained by the requirement of the invariance of the
spacetime line element under an infinitesimal motion along a
vector $\xi^a$;
$$x^a \rightarrow x^a + \delta\lambda \; \xi^a
 \;\;\;\;\;\;\;\; \delta\lambda \ll 1 \eqno(25)$$
Here we use the same mathematical formulation but apply it not to
the spacetime line element $ds^2$  but to the spatial and temporal
line elements $dT^2$ and $dL^2$ of (1). This should be done in
such a way that the freedom in choosing the time zero to be
incorporated in the definition of ({\it physical}) symmetry. To do
so we start from the spacetime line element in form (5);
$$ds^{2}=  h(A_a {\rm d}x^a)^2 - h_{ab}{\rm d}x^a
{\rm d}x^b $$ and we ensure the physical invariance under the
symmetry operation (25) by the following requirements;\\
({\bf a})- $\delta(dL^2)=0$ from which we have;
 $${\pounds_\xi} h_{ab}=0 \eqno(26)$$
({\bf b})- $\delta h =0 $ by which we get;\\
$$\pounds_{\xi}h=0$$
which in turn by (18) reduces to;
$$\xi^a (E_g)_ a = 0$$
and now due to the fact that $(E_g)_ {a ,b}=(E_g)_{ b ,a}$ we
have,
$$\pounds_\xi (E_g)_a =0\eqno(27)$$
({\bf c})- finally to incorporate the freedom in choosing time
zero we need
$$\delta(A_a dx^a) = \delta\lambda \;d \phi \eqno(27a)$$
where $\phi$ is an arbitrary scalar function such that
$\phi(x^a)\doteq \phi(x^\alpha)$, this is so because under the
transformation
$$dx^0 \;\;\;\;\;\; \rightarrow \;\;\;\;\;\; dx^{\prime 0} = dx^0
- \delta\lambda \; d \phi(x^a)$$ we have
$$A_a dx^a \;\; \rightarrow \;\; A_a dx^a +
\delta\lambda \; d \phi \doteq - dx^0 + A_{\alpha} dx^{\alpha} +
\delta\lambda \; d \phi $$ on the other hand under (25)
$$\delta(A_a dx^a)=\delta\lambda \; dx^a \pounds_\xi A_a\eqno(27b)$$
Now comparing equations (27a) and (27b) one gets;
$$\pounds_\xi A_a = \phi_{;a} $$
Using the above relation for the Lie derivative of the
gravitomagnetic potential and the definition of
$F_{ab}=A_{b;a}-A_{a;b}$ one could easily show that;
$$\pounds_\xi F_{ab} = 0 \eqno(28)$$
where $F_{ab}$, in the preferred coordinate system, has the
following form;
$$
F_{ab}\doteq \left(
\begin{array}{cccc}
 0 & 0 & 0 & 0 \\
 0 & 0 & {\hat B}^3 & -{\hat B}^2 \\
 0 & -{\hat B}^3 & 0 & {\hat B}^1 \\
 0 & {\hat B}^2 & -{\hat B}^1 & 0
\end{array}
\right)
$$
or in a more compact, 3-dimensional notation;
$${\hat B}^\alpha =\sqrt{\gamma} B^\alpha = {1\over 2}
\sqrt{\gamma}\varepsilon^{\alpha\beta\gamma} F_{\beta\gamma}$$ One
should note that in the above discussion the appearance of $E_a$
and $F_{ab}$ (and consequently the components $B^\alpha$) is
completely independent of their introduction through the Einstein
field equations in the quasi-Maxwell form. Furthermore we have
shown that the physical symmetry of a spacetime is the one which
is respected by its gravitoelectromagnetic fields and the spatial
metric $h_{ab}$.
\subsection{Gauged motion}
As a consequence of relation (4) and equations (26)-(28), we
obtain,
$$\pounds_\xi g_{ab}=h \phi_{;(a}A_{b)}\eqno(29)$$
This introduces a new generalization of the usual Killing motion
(based on the Killing equation (24)), which we call a {\it gauged
Killing motion} with the corresponding {\it gauged Killing
vector}. \footnote{One should note that there are other
generalization of equation (24) such as the {\it conformal} and
{\it homothetic} Killing motions [12].} It can be shown [13] that
this motion is in accordance with the symmetry of the curvature
invariants of a spacetime and therefore with the spacetime
symmetry itself as it has already been shown for the spherical
[3,10] and cylindrical [11] cases. In the next section we will
apply the above formalism to the special case of plane symmetric
spacetimes.

\section{Plane symmetric spacetimes} Applying the above ideas to
the case of plane symmetry we use the Cartesian coordinates
$x,y,z$ and choose the $z$ direction as the distinguished one for
describing the plane symmetry. Now working in the preferred
coordinate system in which the Killing vectors are given by
$$\xi_{(0)}^a=(1,0,0,0)\;\;\;\;\;\;\;\;\;\;\;\;\;\;\;
\xi_{(1)}^a=(0,1,0,0)$$
$$\xi_{(2)}^a=(0,0,1,0)\;\;\;\;\;\;\;\;\;\;\;\;\;\;\;
\xi_{(3)}^a=(0,-y,x,0)$$ and using equations (26)-(28) one can
show that for a plane symmetric spacetime the following conditions
are required to be satisfied;\\
({\bf a})- from equation (27) and for $\xi_{(1)}^a$ and
$\xi_{(2)}^a$ we find $\nu_{,x}=\nu_{,y}=0$ i.e. $g_{00}$ is only
$z$-dependent. By this condition, on each $(x,y)$ plane, the rates
of the clocks at different points are equal and the
gravitoelectric  field is a constant vector with the only
component ${E_g}_{3}(z)$.\\
({\bf b})- from equation (26) and for $\xi_{(1)}^a$ and
$\xi_{(2)}^a$ one finds,
$\gamma_{\alpha\beta,x}=\gamma_{\alpha\beta,y}=0$ i.e.
components of the spatial metric are only $z$ dependent.\\
({\bf c})- from equation (26) and for $\xi_{(3)}^a$ we find that
on each $(x,y)$ plane, distance elements along the $x$ and $y$
directions are equal and hence we need $\gamma_{xx}=\gamma_{yy}$.\\
({\bf d})- from equation (28) and for $\xi_{(1)}^a$, $\xi_{(2)}^a$
and $\xi_{(3)}^a$ one can show that the gravitomagnetic field
should only be $z$-dependent and that $B_x=B_y=0$.\\
Applying the above conditions and the existence of a timelike
Killing vector $\xi_{(0)}^a$, we obtain the following general form
for a plane symmetric spacetime;
$$ds^2 = e^{2 \nu (z)}(dx^{0}-A_\alpha dx^{\alpha})^{2}-
\gamma_{\alpha\beta}(z) dx^{\alpha}dx^{\beta}$$  On the other hand
using the definition of divergence in $\gamma$-space and the fact
that $B_g^{\alpha}=(0,0,B_g^{3}(z))$, from equation (2) we have;
$$B_g^{3}=\ell/ \sqrt{\gamma}\eqno(30)$$ where
$\gamma={\rm det}{\gamma_{\alpha\beta}}$ and $\ell$ is a constant
indicating the gravitomagnetic field strength. Of course there are
different choices for the vector field $\bf A$  giving rise to the
same ${\bf B}_g$ field but as we have discussed earlier they are
related through gauge transformations induced by a shift in the
time zero. The simplest choices which will do the job are
$A_{\alpha}=(0,\ell x,0)$ and $A_{\alpha}=(-\ell y,0,0)$. Choosing
the first form for the vector potential the metric of a stationary
plane symmetric spacetime with diagonal $\gamma$ matrix will find
the following general form;
$$ds^2= e^{2\nu(z)}(dx^{0}-\ell x dy)^{2}-
e^{\lambda_{\alpha}(z)}(dx^{\alpha})^{2}\eqno(31)$$ where
$\gamma_{\alpha\alpha}=e^{\lambda_{\alpha}}(z)$. For the moment we
forget the condition $\gamma_{11}=\gamma_{22}$ to obtain a more
general solution. Using the form (9) the vacuum equations
$R_{ab}=0$ or their equivalent quasi-Maxwell form lead to the
following equations;
$$2\nu^{\prime \prime}+\nu^{\prime}(2\nu^{\prime}+\lambda_{1}^{\prime}+
\lambda_{2}^{\prime}-\lambda_{3}^{\prime})+\ell^{2}e^{2\nu+
\lambda_{3}-\lambda_{1}-\lambda_{2}}=0\eqno(32)$$
$$-2\lambda_{1}^{\prime \prime}-\lambda_{1}^{\prime}(2\nu^{\prime}+
\lambda_{1}^{\prime}+\lambda_{2}^{\prime}-\lambda_{3}^{\prime})
+2\ell^{2}e^{2\nu+\lambda_{3}-\lambda_{1}-\lambda_{2}}=0\eqno(33)$$
$$-2\lambda_{2}^{\prime \prime}-\lambda_{2}^{\prime}(-2\nu^{\prime}
+\lambda_{1}^{\prime}+\lambda_{2}^{\prime}-\lambda_{3}^{\prime})
+2\ell^{2}e^{2\nu+\lambda_{3}-\lambda_{1}-\lambda_{2}}=0\eqno(34)$$
$$4\nu^{\prime\prime}+4(\nu^{\prime})^{2}-\lambda_{3}^{\prime}(2\nu^{\prime}
+\lambda_{1}^{\prime}+\lambda_{2}^{\prime}) +2\lambda_{1}^{\prime
\prime}+2\lambda_{2}^{\prime \prime}
+(\lambda_{1}^{\prime})^{2}+(\lambda_{2}^{\prime})^{2}=0
\eqno(35)$$ Before looking for the general solution of the above
equations we solve them in two special cases.
\section{THE STATIC CASE}
In this case we set $\ell = 0$ and there are no cross
terms in the metric and consequently ${\bf B}_g=0$.
The resulted equations are easy to solve and the general solution
is given by;
$$g_{00}=e^{2\nu}$$
$$g_{11}=-k_{1}e^{c\nu}$$
$$g_{22}=-k_{2}e^{{-2c\over{c+2}}\nu}$$
$$g_{33}=-k_{3}{\nu^{\prime}}^{2}e^{(2+{c^{2}\over{c+2}})\nu} \eqno(36)$$\\
where $k_\alpha$s and $c$ are constants to be determined. We set
$k_{1}=k_{2}=1$ as they can be absorbed in the coordinates $x_{1}$
and $x_{2}$ respectively. Applying the third condition of plane
symmetry, i.e. $\gamma_{11}=\gamma_{22}$, to the above metric we
get $c=-4$ and so the metric will take the following form;
$$g_{00}=e^{2\nu}$$
$$g_{11}=g_{22}=-e^{-4\nu}$$
$$g_{33}=-k{\nu^{\prime}}^{2}e^{-6\nu}\eqno(37)$$\\
where $k=k_3 > 0$ is the only constant left to be determined. One
should note that this is the same metric found by Taub in his
search for a plane symmetric metric. His metric is given by [8];
$${\rm ds}^2={1\over \sqrt{1+\kappa Z}}({\rm d}t^2-{\rm d}Z^2) -
(1+\kappa Z)({\rm d}x^2 + {\rm d}y^2)\eqno(38)$$ where $\kappa =
constant$. The two forms (38) and (37) can  be transformed into
one another by the following transformation;
$${1\over \sqrt{1+\kappa Z}} = e^{2\nu (z)}$$
with $k={\kappa^2 \over 4}$.
\subsection{A candidate metric for the spacetime of an infintely large massive wall}
Now we would like to find a candidate for the spacetime of an
infinitely large massive wall, whose metric is expected to belong
to the above general plane symmetric static family. But before
looking for a physical potential $\nu(z)$, using the general form
(37), we show that the asymptotic flatness of such a spacetime is
in contradiction with its gravity field being attractive. This is
so because its Riemann invariant is found to be
$$R^{abcd}R_{abcd} = {192\over k^2}e^{12\nu(z)}$$
therefore to get asymptotic flatness we need $\nu(z) \rightarrow
-\infty $ as $z \rightarrow \infty $. On the other hand an
attractive gravitoelectric field for all $z > 0$ requires a
monotonically increasing $\nu(z)$ which is obviously inconsistent
with the previous requirement. This result was not unexpected as
the plane symmetry requirement naturally brings in infinitely
extended sources. An infinitely large massive wall in Newotonian
gravity produces the Newotonian gravitoelectric field ${\bf
E}_g=-{m\over 2} {\hat z}$ for $z\geq 0$, where m is the {\it mass
density} on the wall (the case for $z\leq 0$ is very similar). To
obtain the candidate metric of such a wall we need to introduce a
gravitoelectric potential $\nu(z)$ along with a proper choice of
the constant $k$ such that the resulted metric fulfills the
following two requirements; First of all, in the Newotonian limit,
it should produce the gravitoelectric field ${\bf E}_g=-{m\over 2}
\hat z$ and secondly, for $m=0$,
the metric should reduce to that of the flat spacetime.\\
One can easily see that the choices
$$e^{2\nu(z)} = z^{m^2} e^{m z}\;\;\;\;\;\;\; {\rm and}\;\;\;\;\;\;\;
k=4/m^2 \eqno(39)$$ will do the job and the metric will take the
following form;
$${\rm ds}^2=z^{m^2} e^{m z}{\rm d}t^2- (z^{-2m^2} e^{-2m z})({\rm d}x^2
+ {\rm d}y^2)-(z^{-3m^2} e^{-3m z})(1+m/z)^2{\rm d}z^2 \eqno(40) $$\\
As it can easily be seen this metric is flat for $m=0$ and its
gravitoelectric field $E_g = -{m\over 2}(1+ m/z)$ is different
from that in the Newtonian case but tends to it as $z \rightarrow
\infty $ (though the source is not confined within a limited
region of space, it is localized in the $z$-direction). The
spacetime is not asymptotically flat and the metric components
have nearly the same behaviour at $z=0$ and $z \rightarrow \infty
$ as does Levi Civita's cylindrically symmetric metric [14] at
$\rho=0$ and $\rho \rightarrow \infty $. \footnote {We compare
(40) with Levi Civita's cylindrically symmetric metric because
they are both expected to be the spacetimes of noncompact
sources.}
\section{the stationary case}
The sharpest distinction between the stationary and the static
cases is of course the appearance of the gravitomagnetic field (of
course we need the the cross term to be coordinate dependent), to
which 'nonzero rest mass' particles respond with their velocities
by the second term of the gravitoelectromagnetic force (20).
Normally it is expected that such a field to be produced by  mass
currents but there are some interesting exceptions such as the
NUT-type spacetimes whose gravitomagnetic could not be attributed
to any kind of mass current in a consistent way. These are
stationary solutions of Einstein field equations with two
parameters, the mass parameter $m$ and the so called NUT factor
$\ell$ (also called magnetic mass) which is the source of the
gravitomagnetic field [10,3]. It has already been shown that the
famous {\it spherical} NUT spacetime and its {\it cylindrical
analogue} are the empty space generalizations of the Schwartzchild
and Levi-civita spacetimes respectively. In what follows we will
find the plane symmetric analogue of NUT space and show that it is
the empty space generalization of the metric (40) of the
infinitely large massive wall. But before that we find another
solution of equations (32)-(35) which is a special case of our
final plane symmetric analogue of NUT metric and that is the one
parameter ( $\ell \neq 0$ and $m=0$)  metric corresponding to the
spacetime of a distribution of magnetic masses (NUT parameter)
over an infinite plane, what can be called the plane symmetric
analogue of pure NUT space.
\subsection{PLANE SYMMETRIC ANALOGUE OF PURE NUT SPACETIME}
The following special choice;
 $$ \lambda_{3}=\lambda_{1}+\lambda_{2}+2\nu \eqno(41)$$
cancels out a lot of terms in equations (32)-(35) and what is left
can be easily solved. The only solution which reduces to that of
the Minkowski space when $\ell \rightarrow 0$ is found to be;
$$\nu(z)=-{1\over{2}} ln(cosh(\ell z))$$
$$ \lambda_{1}=ln(cosh(\ell z))+c \ell z $$
$$ \lambda_{2}=ln(cosh(\ell z))+{1\over{c}} \ell z $$
$$ \lambda_{3}=ln(cosh(\ell z))+ (c+{1\over{c}}) \ell z \eqno(42) $$
where c is turned up to be equal to $1$   by the symmetry requirement
$\gamma_{xx}=\gamma_{yy}$  and so the metric will take the following form;
$$g_{00}={1\over{cosh(\ell z)}} $$
$$g_{11}=-cosh(\ell z) e^{\ell z} $$
$$g_{22}=-cosh(\ell z) e^{\ell z}+ {\ell^{2}x^{2}\over{cosh(\ell z)}}$$
$$g_{33}=-cosh(\ell z) e^{2\ell z} $$
$$g_{01}=-{\ell x \over{cosh(\ell z)}} \eqno(43)$$
This is  a stationary spacetime
in which the source of the spacetime curvature is not dependent
on the mass but on another parameter,  the gravitomagnetic field strength $\ell$.
This can also be seen by looking at its gravitomagnetic and
gravitoelecteric fields which are;
$$B_g=\ell {e^{-2\ell z}\over{cosh^{3/2}(\ell z)}}\eqno(44)$$
$$E_g={\ell \over2}tanh(\ell z)\eqno(45)$$
One can compare (43-45) with the usual pure ($m=0$) NUT space
where we have a spherically symmetric stationary spacetime whose
source is the NUT factor $\ell$ and not the mass. Next we find the
promised general family of plane symmetric analogue of NUT
spacetimes where (40) and (43) are its limiting cases for $\ell=0$
and $m=0$ respectively.
\section{PLANE SYMMETRIC ANALOGUE OF NUT SPACE}
To find the general plane symmetric analogue of NUT space (i.e.
$m\neq 0$, $ell \neq 0$) we introduce the following convenient new
variable;
$$2F(z)=\lambda_{3}-\lambda_{1}-\lambda_{2}-2\nu \eqno(46)$$
into equations (32)-(35) which simplifies them and one can show
that their general solution is given by;
$$\lambda_{1}= -2\nu -2c \arctan {\rm h} (\sqrt{1-\ell^{2}d^{2}e^{4\nu}})+\ln (d_1)\eqno(47)$$
$$ \lambda_{2}= -2\nu -{1\over {2c}} \arctan {\rm h}(\sqrt{1-\ell^{2}d^{2}e^{4\nu}})
+ \ln (d_2)\eqno(48)$$
$$ F(z)= \ln(\lambda_{1}^{\prime}+2\nu^{\prime}) + \ln(d) \eqno(49)$$
where coefficients $d1$, $d2$, $d$ and $c$ are constants to be
determined using symmetries and physical arguments. First of all
by the symmetry condition $\gamma_{xx}=\gamma_{yy}$, we find
$c^{2}={1\over{4}}$ and $d_1 =d_2$. Now the requirement that this
solution reduces to that of the general static solution (equation
(37) ) for $\ell=0$ and to that of the flat space for $m=0$ and
$\ell=0$, leads us to the following choices of the constants in
the above functions;
$$d_1=d_2=2^{-m\over m+\ell}{\ell \over (m+\ell)^2}\eqno(50)$$
$$c=-1/2\;\;\;\;\;\;\;\;\;\;\;\;\;\;\; d={1 \over ( m+ \ell)^2}\eqno(51)$$
Now we Choose a suitable  potential $\nu(z)$ so that we could
recover the candidate spacetime of an infinitely large massive
wall when $\ell=0$ and the plane symmetric analogue of pure NUT
space ( equation (43) ) when $m=0$. One can show that the
following choice fulfills these requirements and further we have a
well-defined metric everywhere (for all values of $z$);
$$\nu(z)={1\over{2}} [{Nmz\over {mz+N}} +
ln({z^{m^2}\over cosh(\ell z)})]\eqno(52)$$ where $N = \ln ({m
\over{\ell}}+1)$.
This concludes our general stationary plane
symmetric analogue of NUT space which has two parameter $m$ and
$l$.
\section*{Acknowledgements}
We would like to thank Chris Hillman for his comment on the first
version of the preprint. The authors would like to thank Tehran
University for supporting this project under the grants provided
by the research council.

\pagebreak

\end{document}